# Hall effect in the vicinity of quantum critical point in Tm$_{1-x}$Yb$_x$B$_{12}$.

N.E. Sluchanko[1*], A.N. Azarevich[1], A.V. Bogach[1], V.V. Glushkov[1,2], S.V. Demishev[1,2], A.V. Levchenko[3], V.B. Filipov[3], N.Yu. Shitsevalova[3], J. Stankiewicz[4], V.V. Moshchalkov[5]

[1]-A.M. Prokhorov General Physics Institute, Russian Academy of Sciences, 38, Vavilova str., 119991, Moscow, Russia

[2]-Moscow Institute of Physics and Technology, 9, Institutskii per., 141700, Dolgoprudnyi Moscow Region, Russia

[3]-Institute for Materials Science Problems, National Academy of Sciences of Ukraine, 3, Krzhyzhanovsky str., 03680, Kiev, Ukraine

[4]- Institute for Materials Science of Aragón, 50009 Zaragoza, Spain

[5]- Institute for Nanoscale Physics and Chemistry, KULeuven, Celestijnenlaan 200 D, B-3001 Leuven, Belgium

*-E-mail: nes@lt.gpi.ru

The angular, temperature and magnetic field dependences of Hall resistance $\rho_H$ for the rare-earth dodecaboride solid solutions Tm$_{1-x}$Yb$_x$B$_{12}$ have been studied in a wide vicinity of the quantum critical point (QCP) $x_C \sim 0.3$. The measurements performed in the temperature range 1.9÷300 K on high quality single crystals allowed to find out for the first time in these *fcc* compounds both an appearance of the second harmonic contribution in $\rho_H$ at QCP and its' enhancement under the Tm to ytterbium substitution and/or with increase of external magnetic field. When the Yb concentration $x$ increases a negative maximum of a significant amplitude was shown to appear on the temperature dependences of Hall coefficient $R_H(T)$ for the Tm$_{1-x}$Yb$_x$B$_{12}$ compounds. Moreover, a complicated activation type behavior of the Hall coefficient is observed at intermediate temperatures for $x \geq 0.5$ with activation energies $E_g \sim 200K$ and $E_a \sim 55$-$75K$ in combination with the sign inversion of $R_H(T)$ at low temperatures in the coherent regime. The density of states renormalization effects are analyzed within the variation of Yb concentration and the features of the charge transport in various regimes (charge gap formation, intra-gap manybody resonance and coherent regime) are discussed in detail in $Tm_{1-x}Yb_xB_{12}$ solid solutions.

# 1. INTRODUCTION

The Hall effect measurements are considered as one of the key experiments to reveal a transformation of electron excitation spectrum in a vicinity of quantum critical point (QCP) [1-2]. Rather recently, anomalies of Hall coefficient were observed near an antiferromagnetic (AF) QCP (Neel temperature $T_N=0$) which corresponds to the transition from AF to paramagnetic (P) state in the heavy fermion (HF) systems with metallic conductivity [3-6]. Since the magnetic order in HF metals is usually attributed to the appearance of a magnetic structure of localized magnetic moments (LMMs) of the rare-earth (RE) ions, the given AF-P transition from magnetic to nonmagnetic HF ground state through the QCP may be accompanied simultaneously by delocalization of the RE ions' electronic states.

In the ongoing discussion on the nature of the electron structure transformation in the RE compounds in the nearest vicinity of QCP, two possible scenarios for realization of the quantum phase transition are the most actively analyzed. According to the first approach, at the QCP the RE localized magnetic moments (LMM) are coupled with itinerant electrons in the manybody states, so that these heavy fermions form antinodes of spin-density wave (SDW) in the metallic matrix (see Refs. 7-9). In the alternative scenario an abrupt change of the Fermi surface's volume occurs at the QCP owing to the decay of the manybody states and it causes an "unfreezing" of the LMMs of the rare-earth ions [1,9]. A quite different, the third one, approach to the interpretation of the charge transport anomalies in substitutional solid solutions $CeCu_{6-x}Au_x$ was proposed rather recently in Ref.6, where the Hall effect has been studied in detail in a wide vicinity of AF quantum critical point $x_C \approx 0.1$. In particular, for along with a remarkable increase of Hall coefficient near the QCP, the appearance of even (in magnetic field) harmonics in the Hall resistance observed in [6] was attributed to the occurrence of an electron type phase segregation in these HF compounds in a vicinity of QCP. Taking in mind the examples of the systems with fast *4f-5d* spin fluctuations and AF quantum critical point in which the metal-insulator transition (MIT) and the AF-P transition occur simultaneously, it is of interest to investigate the Hall effect features in a vicinity of the AF quantum critical point with MIT.

Among the most convenient compounds for these test experiments is the system $Tm_{1-x}Yb_xB_{12}$, where the thulium to ytterbium substitution causes to a development of AF-instability approaching to QCP at $x = x_C \approx 0.3$ [10] and the metal-insulator transition is observed in a wide vicinity of the quantum critical point. It is worth noting here that according to the results of Seebeck coefficient measurements [10] in the range $x \leq 0.3$ the activation energy $E_g/2$ on the magnetic side of the AF-P transition increases drastically from zero in the antiferromagnetic metal $TmB_{12}$. Above $x_C$ the deduced value $E_g/k_B = 120 \div 160$ K [10] becomes equal approximately to the band gap obtained from the charge transport [11], NMR [12] and the heat capacity measurements [13] for $YbB_{12}$ narrow gap semiconductor with strong valence fluctuations. Thus, the main goal of this research was to carry out a precise Hall effect studies of the $Tm_{1-x}Yb_xB_{12}$ compounds with AF quantum critical point and metal-insulator transition.

## 2. EXPERIMENTAL DETAILS

In the present study, the angular, temperature and magnetic field dependences of the Hall resistance $\rho_H(\varphi, H_0, T_0)$ and resistivity $\rho(T)$ were investigated for substitutional solid solutions $Tm_{1-x}Yb_xB_{12}$ in the range of compositions $0 \leq x < 0.75$ at temperatures in the interval $1.9 \div 300$ K in magnetic field up to 8 T. The measurements were performed on high-quality single crystals of the rare-earth dodecaborides. The samples were grown in Kiev (IPMS NASU) by vertical crucibleless inductive zone melting in the inert gas atmosphere as described in [14]. To study the Hall resistance the original sample rotation technique was applied with a stepwise fixing of the sample's position in the steady magnetic field which was transversal to the DC measuring current $I \parallel <110>$ (see an inset on Fig. 1a), the installation is similar to that one applied earlier in [6,15]. A special attention was paid to the symmetric location of the Hall probes to minimize the contribution from the negative magnetoresistance. It should be stressed that the transverse configuration applied here for the Hall effect measurements (the axis of the sample rotation is parallel to the direction of the measuring current *I* and perpendicular to *H*; see the schema in the inset in Fig. 1) eliminates the contribution of the longitudinal magnetoresistance to the Hall signal [6,15].

### 3. RESULTS AND DISCUSSION

**3.1 Resistivity.** The results of the resistivity $\rho(T)$ measurements on the $Tm_{1-x}Yb_xB_{12}$ samples with a various Yb content are presented in fig. 1. It is visible from the data in Fig. 1 that the $\rho(T)$ dependence for antiferromagnet $TmB_{12}$ ($x=0$, $T_N=3.2$ K) performs a typical metallic behavior with both a significant resistivity decrease with the temperature lowering and small enough residual values $\rho_0 \sim 4$ μΩ·cm. Thulium to ytterbium substitution in $Tm_{1-x}Yb_xB_{12}$ solid solutions with $x<x_C$ is accompanied by both an appearance of a pronounced resistivity minimum at temperatures above 12 K and a strong enhancement of the $\rho(T)$ spin-flip scattering contribution with elevation of Yb content. Analysis of the resistivity temperature dependences in the wide vicinity of QCP (fig. 1) allows concluding in favor of the metal–insulator transition in the $Tm_{1-x}Yb_xB_{12}$ dodecaborides. Indeed, when $x$ varies in the range $0 \leq x < 0.75$, the resistivity at helium temperatures increases by more than two orders of magnitude, and the increase cannot be interpreted in terms of the charge carriers' scattering caused by the substitution disorder. Meanwhile, it should be mentioned here that the activation behavior of $\rho(T)$ is not observed throughout the concentration range $x < 0.75$.

**3.2. Hall resistivity contributions.** The Hall resistivity angular dependences are shown in figs. 2, 3 as the typical examples obtained both at various temperatures for $Tm_{0.46}Yb_{0.54}B_{12}$ in the magnetic field $\mu_0H=1.5$ T and at helium temperature $T_0=2.1$ K for several magnetic field values in the range $\mu_0H \leq 8$T for $Tm_{0.28}Yb_{0.72}B_{12}$ respectively. It should be mentioned that except for the antiferromagnet $TmB_{12}$ ($T_N=3.2$), all the $Tm_{1-x}Yb_xB_{12}$ samples with Yb content in the interval $0.1<x<0.75$ investigated here are paramagnets in a range of temperatures $T>1.9$ K. Moreover, as the dodecaborides are crystallized in the *fcc* structure, it is reasonable to expect a cosine type angular dependence of the Hall resistivity $\rho_H(\varphi) \sim \rho_{H1} \cos\varphi$ when the sample is rotating in magnetic field (see inset in fig. 1a). Such a behavior of $\rho_H(\varphi)$ was observed in our experiments for the solid solutions $Tm_{1-x}Yb_xB_{12}$ in the range $0.1<x<0.6$ at temperatures above 30 K and in magnetic fields up to 2 T (see, for example, fig. 2a) and it can be attributed to a variation of the ***H*** vector's normal component in according with the harmonic law when the sample is rotating in magnetic field. Upon the decrease of temperature in the range of $T<30$ K a remarkable contribution from the second harmonic appears in the Hall signal (see figs. 2b-d, 3a-d). In this study we have analyzed the angular dependence of the Hall resistivity $\rho_H(\varphi, T_0, H_0)$ of $Tm_{1-x}Yb_xB_{12}$ solid solutions by the relation

$$\rho_H(\varphi, T_0, H_0) = \rho_{H0} + \rho_{H1} \cos\varphi + \rho_{H2} \cos(2\varphi - \Delta\varphi) \qquad (1),$$

where in addition to the main contribution $\rho_{H1}$ and to the constant bias $\rho_{H0}$ the second harmonic's term $\rho_{H2}$ which is even in magnetic field, is also included and it is taken into account with a phase shift $\Delta\varphi$ with respect to the main component of the Hall signal. It is discerned in figs. 2-3 that at liquid helium temperatures for $Tm_{1-x}Yb_xB_{12}$ compounds with Yb concentration $x$ above QCP $x_C \approx 0.3$, the amplitude $\rho_{H2}$ of this even component in the Hall effect becomes comparable or exceeds considerably the usual odd cosine contribution to the Hall resistivity $\rho_H(\varphi)$ in the magnetic fields $1T<\mu_0H<3T$. Then, the odd component's sign inversion is observed for $x \geq 0.5$ both with the temperature lowering in the range 4.2-1.9 K (see for comparison figs. 2c and 2d) and when the magnetic field increases at $T_0=2.1$ K (figs. 3b and 3c). Moreover, because of the small but finite phase shift $\Delta\varphi \neq 0$ between the main $\rho_{H1}$ and even $\rho_{H2}$ components, the position of the extremuma on the experimental curves $\rho_H(\varphi, T_0, H_0)$ disagrees with two normal vector ***n*** directions (see inset in fig.1, $\varphi=0°$ and $\varphi=180°$).

In such a situation the conventional technique usually employed for Hall effect studies is not appropriate for the separation and analysis of the Hall resistivity components. Indeed, the conventional technique is based on the measurement of the Hall signal at two values $\varphi=0°$ and 180° of the angle between $\mathbf{n}$ and $\mathbf{H}$ vectors with estimation of the Hall resistivity as the half-difference *[$\rho_H(\varphi=0)–\rho_H(\varphi=180°)$] /2* resulting to a loss of this even harmonic signal in the final product. Moreover, it leads evidently to incorrect evaluation of the Hall coefficient for $Tm_{1-x}Yb_xB_{12}$ compounds with Yb content $x>x_C$ in the case of any small enough non-colinearity of the $\mathbf{n}$ and $\mathbf{H}$ vectors (see, for example, figs.2-3). Thus, the measurements of the $\rho_H(\varphi,T_0,H_0)$ angular dependences and their analysis in the framework of the relation (1) is of crucial importance and it is the only one correct technique to analyze the Hall effect contributions in the $Tm_{1-x}Yb_xB_{12}$ system with quantum critical point $x_C\approx0.3$ and MIT.

To study the Hall effect features in $Tm_{1-x}Yb_xB_{12}$ in a wide vicinity of QCP and under the metal-insulator transition we performed detailed investigations of the angular dependences $\rho_H(\varphi)$ for compounds with $x\approx0.31\approx x_C$ (QCP), $x\approx0.54$ and $x\approx0.72$ at temperatures in the range 1.9-300K and in magnetic field up to 8 T. Further on, these main (odd) and even Hall effect contributions detected in according with the relation (1) have been analyzed quantitatively. An examples of the analysis of the $\rho_H(\varphi)$ curves are presented on figs. 2b-d and 3a-d, where the decomposition of the experimental dependences on the main and even Hall resistivity components are shown at liquid helium temperatures for various magnetic fields below $\mu_0H=8$ T (see dashed and dotted curves in figs.2-3). Apparently from figs. 2-3, the relation (1) describes accurately the experimental data (solid lines in figs.2-3 perform the sum of these two components in $\rho_H(\varphi)$) and allows both to define the amplitude values $\rho_{H1}(H,T)$ and $\rho_{H2}(H,T)$ for all compounds investigated and to estimate the phase shift $\Delta\varphi(T,H)$ between the components in the Hall signal. Thus, within an experimental accuracy the phase shift serves as a constant $\Delta\varphi(T,H)\approx15°$, whereas the amplitudes $\rho_{H1}$ and $\rho_{H2}$ for $Tm_{1-x}Yb_xB_{12}$ compounds vary considerably as a function of temperature and magnetic field. The temperature dependences of the Hall coefficient $R_H(H_0,T)=\rho_{H1}(H_0,T)/H$ and the second harmonic's amplitude $\rho_{H2}(H_0,T)$ are shown in fig.4 and fig.5, respectively. Magnetic field dependences of Hall coefficient $R_H(T_0,H)$ and the Hall resistivity even component $\rho_{H2}(T_0,H)$ obtained from the measurements at liquid helium temperatures ($T_0$=2.1K and 4.2K) for different ytterbium concentrations are presented in figs. 6 and 7, correspondingly. For comparison, the results of the Hall coefficient measurements for $TmB_{12}$ ($x$=0) in the small magnetic field $\mu_0H$=0.37 T are also shown in fig. 4.

### 3.3. Analysis of the contributions to the Hall effect.

**3.3.1. Hall coefficient.** The usual (odd in magnetic field) contribution to the Hall effect increases drastically for Yb contents above quantum critical point ($x \geq x_C \approx 0.3$) with the temperature lowering in the range 30-300 K. Below 30K the amplitude of the negative Hall coefficient is strongly depressed and the $R_H(T)$ sign inversion is observed at liquid helium temperatures in the small magnetic field $\mu_0H \leq 2$ T for the samples with Yb contents $x > x_C$ (figs.4,6). As a result, the Hall coefficient of $Tm_{1-x}Yb_xB_{12}$ system with $x \geq x_C$ reveals a large amplitude negative maximum at intermediate temperatures 10-50 K and when $x$ increases the metal-insulator transition is accompanied with a shifting of the remarkable feature of $R_H(T)$ downwards on the temperature scale together with a simultaneous elevation of its amplitude.

Similar behavior of an opposite (positive) sign Hall coefficient with a maximum $R_H(T)$ and the sign inversion at low temperatures was observed earlier for dense Ce- and U-based metallic heavy fermion systems $CeAl_3$ [16], $CeCu_6$ [17], $CePd_3$ [18], $UAl_2$ [19], $UPt_3$ [20], and for so-called Kondo-insulator $CeNiSn$ [21]. The effect was attributed to the appearance of a many-body resonance at Fermi energy $E_F$ in the electron density of states (DOS) and to a crossover to a coherent regime of a magnetic scattering of the charge carriers in these strongly correlated electron systems. Moreover, with the help of detailed Hall effect measurements of the archetypal heavy fermion compounds $CeAl_2$ [15], $CeAl_3$ [22,6], $CeCu_6$ [6] and $CeB_6$ [23] a complicated activation type dependence of the Hall coefficient

$$R_H(T) \sim \exp(E_{a1,2}/k_BT) \qquad (2)$$

was certainly established in the temperature range $T>T^{RH}_{max}$. It was noted previously [6,15,22-23], that the activation type behavior of $R_H(T)$ observed in these Ce-based metals is extremely unusual and does not find explanation in the framework of the Kondo-lattice model or the model of skew-scattering of charge carriers [24-25]. Indeed, in these models the spin-flip resonant scattering of conduction electrons on localized magnetic moments of the rare-earth ions is considered as a dominant factor determining charge transport. Within the approach [24-25] both the anomalous Hall effect and the non-monotonous behavior of resistivity in heavy-fermion compounds should be exclusively attributed to the specific character of the scattering effects. At the same time, from the standpoint of the spin-polaron approach, which is an alternative to the skew-scattering model, the parameters $E_{a1,2}$ found for heavy fermion Ce-based compounds, should be considered as the bound energies of different type many-body states formed in the vicinity of Ce 4$f$ centers [6,15,22-23].

In the case of the Yb-based compounds under investigation, one should take into account that (i) Yb is the hole-type analogue to the Ce and U in these strongly correlated electron systems and (ii) the metal-insulator transition occurs in the $Tm_{1-x}Yb_xB_{12}$ solid solutions. So, the negative $R_H(T)$ extremum (fig. 4) should be also analyzed above the maximum at $T^{RH}_{max}$ in terms of Eq. (2). It is visible from the activation plot presented in fig. 8a that $R_H(T)$ curves for $Tm_{1-x}Yb_xB_{12}$ solid solutions with $x \geq 0.5$ can be well approximated by the activation law (2) in the temperature intervals 120-300K (I) and 50-120K (II). Furthermore in the range I the energy gap $E_g \approx 206$ K $\approx$ 17.8 meV may be determined. The value $E_g \approx 17.8$ meV estimated for $x$=0.54 and 0.72 (fig. 8a) appears to be close to the gap found for narrow gap semiconductor $YbB_{12}$ in the measurements of the Hall effect and resistivity ($E_g/k_B \approx$ 180 and 134 K, respectively [11]), NMR on the Yb ions [12] and specific heat [13] ($E_g/k_B \approx$ 170 K). It is also comparable to the spin gap value ≈ 12 meV found in the inelastic neutron scattering [26-27] and ESR [28] experiments.

Additionally, in the framework of the relation for activation type dependence of Seebeck coefficient

$$S = \frac{k_B}{e} \frac{b-1}{b+1} \frac{E_g}{2k_BT} \qquad (3)$$

where $b = \mu_n/\mu_p$ is the electron/hole mobility ratio and $e$ is the charge of electron, one can deduce also the ratio $\mu_n/\mu_p \sim 7.7$ from the comparison of the slopes of $S \sim 1/T$ [10] and $\ln|R_H| \sim 1/T$ dependences for $Tm_{1-x}Yb_xB_{12}$ solid solutions with $x \geq 0.5$ (fig.8a). It should be emphasized that the ratio $b = \mu_n/\mu_p \approx 50$ was previously obtained for another archetypal narrow gap semiconductor-intermediate valence compound $SmB_6$ [29], and it was shown that the large $b$ values in the regime of fast charge fluctuations are due to the low mobility $\mu_p$ of the heavy holes on the rare earth ions compared to $\mu_n$ for conduction electrons in the $5d$ band.

In the spin-polaron approach the $E_g = 2E_{a1}$ and $E_a \equiv E_{a2}$ parameters in Eq.(2) obtained for $Tm_{1-x}Yb_xB_{12}$ with $x \geq 0.5$ can be treated as bound energies of the short radius many-body states which are formed in the vicinity of Yb $4f$-centers in the matrix of these Yb-based dodecaborides. It is worth noting also that in the charge transport regime via the states in the gap (range II in fig.8a) the activation energy values $E_a$ of the Hall (fig.8a) and Seebeck coefficients are coincide (see also Ref. 10). In the developed approach the value of the width of the quasi-elastic peak $\Gamma_0(T>100K)/2 \approx 12.5$ meV obtained from the neutron scattering experiments in $YbB_{12}$ and $Yb_{1-x}Lu_xB_{12}$ solid solutions [30-31] can be used within a simple relation

$$\Gamma_0(T>100K)/2 = \hbar/\tau_{eff}(T) \qquad (4)$$

to evaluate the relaxation time $\tau_{eff}(T \geq 100K) \approx 1.1 \cdot 10^{-13}$ sec and, further on, to estimate the value of effective mass $m^*$ of the heavy-fermions from Hall mobility $\mu_H(T) = R_H(T)/\rho(T)$ [6,15,22-23]:

$$m^*(T) = e\tau_{eff}(T)/\mu_H(T) \qquad (5).$$

The Hall mobility temperature dependences for $Tm_{1-x}Yb_xB_{12}$ solid solutions are shown in fig.9 as deduced from the data of fig.1 and fig.4. The calculated effective mass $m^*(T \approx 100K) \approx 24\, m_0$ for these Yb-based compounds is consistent with the values determined previously for the effective mass of the spin-polaron and exciton-polaron many-body states in the strongly correlated electron systems: FeSi ($m^* \approx 20–90\, m_0$)[32], $SmB_6$ ($m^* \approx 20–40\, m_0$)[29], $CeAl_2$ ($m_{1,2}^* \approx 55–90\, m_0$)[15], $CeAl_3$ ($m_{1,2}^* \approx 45–90\, m_0$)[22], $CeCu_6$ ($m_{1,2}^* \approx 130–150\, m_0$)[6], $CeB_6$ ($m^* \approx 400\, m_0$)[23] and is also comparable with the estimate $m^*\approx12$–$34\ m_0$ obtained for $YbB_{12}$ from optical measurements[33-34] in the temperature range 20-200 K. Taking into account the known parameters $E_a$, $E_g$ (fig.8a) and effective masses $m^*_{I,II}$ with the use of the relation

$$a^*_{p1,2}= \hbar /(2E_{g,a}m^*_{I,II})^{1/2} \qquad (6)$$

one can also estimate the localization radius of the manybody states. For these heavy fermions in the $Tm_{1-x}Yb_xB_{12}$ matrix in the temperature intervals *I* (120-300K) and *II* (50-120K) this gives the values 5Å and 9Å, respectively. The same size ~5 Å of the many-body exciton-polaron states was obtained in Ref. 33 from far infra-red quasi-optical measurements of $YbB_{12}$. It should be mentioned here that in the temperature range *I* of the gap formation the spatial extension of the complexes $a^*_{p1}\approx5$ Å is about equal to the shortest distance (~5.3 Å) between the rare earth $RE^{3+}$ ions in the $RB_{12}$ lattice; therefore, the many-body states are localized in the unit cell in the vicinity of the RE-sites. With temperature lowering in the range II and approaching to the coherent regime, the localization radius $a^*_{p2}\approx9$ Å starts to exceed the lattice constant ($a$~7.5 Å) and, moreover, $a^*_{p2}$ value increases with elevation of $E_a$ (see fig.8a and Eq.(6)) and simultaneously with Yb concentration *x*. In this respect it is natural to assume that the essential increase of the localization radius induces a transition to the coherent state, which is displayed in the abrupt decrease of Hall (fig. 4) and Seebeck [10] coefficients in the $Tm_{1-x}Yb_xB_{12}$ solid solutions with strong electron correlations.

Another novel experimental result of the present study can be found at temperatures below the negative maximum of Hall coefficient (range III, $T<T^{RH}_{max}$, fig.8b) in $Tm_{1-x}Yb_xB_{12}$ with $x>x_C$ where the temperature variation of $R_H$ follows to the dependence

$$R_H(T)\sim exp(-T_0/T) \qquad (7)$$

with the estimated values of $T_0\approx3.5$-$7K$ (fig.8b). Similar behavior of $R_H(T)$ was observed previously for the heavy fermion compounds $CeAl_2$ [15] and $CeAl_3$ [22] and it was interpreted in terms of the dependence predicted in [35] for Hall coefficient in a system with a Berry phase effects where the carrier motion is hopping in a topologically nontrivial spin background.

According to [35-36], the Hall effect is modified in such a situation because of the appearance of the internal magnetic field $H_{int}= <h_z> \sim (1/k_BT) \, exp(-T_0/T)$, which adds to the external field $H$.

An external magnetic field when inducing a polarization both of the localized magnetic moments of the rare earth ions and the spins of conduction electrons, suppresses the coherence in the spin-flip scattering in the periodic system of the RE centers. As a result, sign inversion of $R_H(H,T_0)$ is observed in magnetic field at helium temperatures (see curves $T_0$=2.1 K in fig. 6 for $Tm_{1-x}Yb_xB_{12}$ with $x$=0.54 and 0.72). Further on, the Hall coefficient $R_H(T)$ is happened to be nearly temperature independent in strong magnetic fields (fig. 4). It should be mentioned that the Hall coefficient behavior in magnetic field in $Tm_{1-x}Yb_xB_{12}$ is quite different from that one reported previously for archetypal Ce- and U-based heavy fermion systems[6,15-22] and observed in $HoB_{12}$ [37]. Indeed, in the mentioned compounds the depression of the Hall effect amplitude is usually observed in high magnetic fields. Moreover, the $R_H(H)$ decrease is much more pronounced at temperatures below the Hall coefficient maximum, in the interval $T<T^{RH}_{max}$. The established here an opposite tendency of the Hall effect enhancement in magnetic field (see fig. 4 for $Tm_{1-x}Yb_xB_{12}$ compounds under investigation) was observed recently for RE-hexaborides- $Ce_{1-x}La_xB_6$ [38], $PrB_6$ and $NdB_6$ [39-40] and it was attributed to the magnetization of the spin-polaron manybody states by external magnetic field.

To conclude this Section, let us roughly estimate the localization radius of the many-body states in another way, from the results presented in fig. 8b. The decrease of the reduced concentration of charge carriers per RE- ion $\nu=(R_H e N_{4f})^{-1}$ (where $N_{4f}$=0.96·10$^{22}$ cm$^{-3}$ is the RE-ions' concentration in $RB_{12}$, R-Tm,Yb) in the interval $0.08<\nu\leq 0.8$ can be treated as an increase in the effective volume per conduction electron, which we believe to be caused by the many-body states' formation in $Tm_{1-x}Yb_xB_{12}$. Taking into account the $R^{+3}-R^{+3}$ distance (~5.3 Å) in the *fcc* crystal structure of the dodecaboride compounds under investigation we obtain the crude estimate $a_{p2}$= 6–12 Å at temperatures in the vicinity of the negative $R_H(T)$ maximum for the solid solutions with Yb-content in the range $x \geq x_C$. It is worth noting that the deduced $a_{p2}$ values agree very well with the aforementioned results obtained on the basement of Eq. (6).

**3.3.2. Second harmonic contribution to the Hall effect.** When analyzing the second harmonic component in the Hall effect of $Tm_{1-x}Yb_xB_{12}$ system it is worth noting that this kind contribution has been found earlier in the strongly correlated antiferromagnets $CeAl_2$[15], $CeB_6$[23,41-42], $PrB_6$[40], $NdB_6$[39-40] and $ErB_{12}$[43] both in the paramagnetic and in the magnetically ordered phases. An even harmonic in the angular dependences of Hall signal was observed also in the narrow-gap semiconductor FeSi[32,44] in the wide range of temperatures near the mictomagnetic phase transition. The detailed study of the second harmonic term was undertaken recently [6,45] for $CeCu_{1-x}Au_x$ solid solutions where the Hall effect anomalies were investigated in vicinity of AF quantum critical point $x_C \approx 0.1$. It was established certainly that the development of antiferromagnetic instability in these model compounds is accompanied both with a remarkable increase of the low temperature Hall coefficient $R_H(x)$ at $x \approx x_C$ and an appearance of even harmonic contribution $\rho_{H2}(x,T,H)$ whose amplitude takes on its' maximum value directly at the QCP [6]. Moreover, for the series of compounds $CeCu_{1-x}Au_x$ in the concentration range $0 \leq x \leq 0.15$ near the QCP the temperature dependencies $\rho_{H2}(T)$ have been approximated with a good accuracy by a hyperbolic dependence of the form

$$\rho_{H2}(T) \approx C^*(1/T - 1/T^*) \qquad (8).$$

The hyperbolic divergence of the Hall resistivity $\rho_{H2}(T)$ was established to be the feature of the regime of quantum critical behavior, together with another anomalies including the logarithmic divergence of the electron term in the heat capacity $C(T)/T \sim \ln T$, non-Fermi-liquid behavior of the resistivity $\rho(T) \sim T$, power-law asymptotic of the magnetic susceptibility $\chi(T) \sim T^{-\beta}$, etc.[46-56], which are associated with divergence of the effective mass of manybody states (heavy fermions) nearby QCP.

The second harmonic contribution in the Hall signal of $Tm_{1-x}Yb_xB_{12}$ solid solutions detected in present study (see fig.5a) was analyzed in magnetic fields below 2 T in terms of Eq. (8). It was found evidently that the $\rho_{H2}(T)$ follows the hyperbolic dependence Eq.(8) with a characteristic temperature $T^*\approx 70K$ which is common for all the Yb-contents in the range $x\geq x_C$. The slope $C^*(x)$ of the hyperbolic dependence increases dramatically (more than 10 times) when the metal-insulator transition develops in the series of $Tm_{1-x}Yb_xB_{12}$ compounds and the tendency to saturation of the $\rho_{H2}(1/T)$ curve is observed at temperatures below 10K only for highest Yb composition (fig.5a). Additionally, the amplitude $\rho_{H2}(T_0,H)$ of this even component in the Hall signal reveals at helium temperatures more than order of magnitude increase in the external magnetic field up to 8 T (fig. 7). The double-logarithmic plot of the Hall resistivity temperature dependence $\rho_{H2}(T,H_0)$ is shown in fig. 5b for highest magnetic fields $\mu_0H_0$=7-8 T applied in the study. In the high magnetic fields the second harmonic contribution appears at $T_{IR}$~130-150K, it increases drastically with the temperature lowering and a saturation of this even component is observed at helium temperatures for all the Yb concentrations $x\geq x_C$ (fig.5b).

At the end of this Section let us summarize both a significant difference and similarities in behavior of the second harmonic contribution near the AF QCP in the heavy fermion systems with metal conductivity, from one hand, and in compounds with MIT, on the other. So, in the heavy fermion metals $CeCu_{1-x}Au_x$ an even anomalous contribution in Hall effect is observed in narrow vicinity of AF QCP [6], whereas in the second type systems with MIT the $\rho_{H2}(H,T)$ component appears close to QCP at $x=x_C$~0.3 and it's amplitude increases dramatically with elevation of $x$ when the metal-insulator transition develops (see figs. 5 and 7). At the same time, in both the metallic and dielectric systems with AF QCP the characteristic temperature $T^*$ in hyperbolic relation (8) is positive (see fig. 5a and [6]) and, in our opinion, $T^*$ values may be likely associated with an appearance of a structural distortions in these compounds. Indeed, the phase transition at $T^*\approx$ 60-70K was found very recently in the reference non-magnetic dodecaboride compound $LuB_{12}$[57]. Moreover, a comprehensive study of the heat capacity and Raman scattering spectra in the $Lu^NB_{12}$ crystals with a various boron isotope concentration (N- 10, 11 and natural) allowed the authors [58] to conclude in favor of the cage-glass phase transition at $T^*$ in the family of the RE-dodecaborides. Besides it was shown in [58] that the combination of loosely bounded state of the RE-ions in the rigid boron sub-lattice of $RB_{12}$ compounds together with the randomly arranged boron vacancies provokes a development of vibrational instability at intermediate temperatures and the Ioffe-Regel limit is reached in the $RB_{12}$ family at $T_{IR} \approx 130$-$150K$. As a result, the anomalies at 150 K both in the μSR-spectra[59-60] and high-resolution PES[61] data together with the Hall mobility features at $T_{IR} \sim 130$-$150K$ (see fig.9) may be attributed to the crossover from a phonon-assisted to a quasi-local vibrations' regime in the charge carriers' scattering. Additionally we need to note that the disorder effects may be enhanced in the vicinity of the quantum critical points[62]. When comparing the metallic and dielectric systems with antiferromagnetic QCP it should be stressed here, that similarly to the structural changes at $T^*$ in $Tm_{1-x}Yb_xB_{12}$, the transition from the orthorhombic to the monoclinic structure was established evidently in QCP solid solution $CeCu_{5.9}Au_{0.1}$ at $T_S \approx 70$ K [63-64], and it was shown that the transformation is accompanied by a slight distortions of the bond angles. Since the regime of quantum critical behavior is suppressed by an external magnetic field the natural explanation may be found for the magnetic field dependences $\rho_{H2}(H)$. Indeed, in both type systems the sharp growth of $\rho_{H2}(H)$ in the range $\mu_0H \leq 3$ T is accompanied with the destroying of the hyperbolic temperature dependence (Eq.8) and, as a result, the regime of quantum critical behavior falls down and a tendency to saturation of $\rho_{H2}(H)$ is observed in magnetic fields above 5 T.

**3.3.3. Quantum critical point and metal-insulator transition.** When discussing the DOS renormalization effects within the metal-insulator transition in the series of $Tm_{1-x}Yb_xB_{12}$ compounds it is of interest to start on from the properties of the metal $TmB_{12}$. Nearly temperature independent behavior of the Hall coefficient was detected for this compound in magnetic fields up to 8 T (see, for example, curve for $\mu_0H=0.37$ T in fig. 4). The value $R_H(T) \approx const \approx -3.25 \cdot 10^{-4}$

cm$^3$/C in the approximation of one group of charge carriers is equal to the reduced carriers' concentration per RE- ion $\nu = (R_H e N_{4f})^{-1} \approx 2$. So, TmB$_{12}$ appears to correspond to the two-electron metal with a good accuracy. Above the Neel temperature $T_N \approx 3.2$ K the resistivity of TmB$_{12}$ exhibits a minimum at 11K and maximum at 6.5 K [65]. Moreover, a remarkable negative magnetoresistance [65] and positive contribution to the Seebeck coefficient [43] at temperatures below 40K have been found. The authors [43,65] concluded in favor of strong local *4f-5d* spin fluctuations and the spin-polaron approach was applied to explain the properties of this dodecaboride.

The Tm to ytterbium substitution in the range $x \leq x_C \sim 0.3$ is accompanied with the development of AF-instability in the Tm$_{1-x}$Yb$_x$B$_{12}$ family resulting to a gradual $T_N$ decrease (fig.10a, the range $0 \leq x \leq 1$ corresponds to filling of the *4f*-shell in the interval $12 \leq n_{4f} \leq 13$ in the Tm$_{1-x}$Yb$_x$B$_{12}$ solid solutions, and $n_{4f}=$ 10, 11 and 14 meet to HoB$_{12}$, ErB$_{12}$ and LuB$_{12}$, respectively). The approximation of $T_N(x)$ dependence to value $T_N=0$ gives $x_C \sim 0.3$ in the series of Tm$_{1-x}$Yb$_x$B$_{12}$ solid solutions (QCP in fig.10a). The regime of quantum critical behavior was likely observed in Tm$_{0.74}$Yb$_{0.26}$B$_{12}$ for magnetic contribution to the specific heat with a logarithmic divergence of the form $C/T \sim \ln T$ at $T<4K$ [66]. At the same time, in [10] the complicated activation type behavior of the Seebeck coefficient *S(T)* (see Eq.3) was found in a range of intermediate temperatures (intervals I and II) for all Tm$_{1-x}$Yb$_x$B$_{12}$ compounds with $x \geq 0.05$. Moreover, for AF compounds with Yb content $0.05 \leq x < x_C \sim 0.3$ both the gap formation in the range I (120-300 K) and the regime of charge transport through the manybody intra-gap states with the bound energy $E_a$ (interval II, 50-120 K) are detected on the *S(T)* dependences simultaneously with a weak localization behavior of resistivity (fig.1) and a moderate growth of Hall coefficient (figs. 4 and 8a). In the paramagnetic state just above the QCP $x_C \sim 0.3$ the increase of Yb concentration induces a remarkable rise of Hall coefficient $R_H(x)$ together with an appearance and a dramatic increase of the second harmonic component $\rho_{H2}(x)$ in the Hall effect (fig.10b). Further on, for $x>0.5$ the activation type behavior of $R_H(T)$ allows one to deduce from the temperature dependence of the reduced carriers' concentration (see fig.8a) the gap value $E_g \approx 17.8$ meV and the width $E_a/k_B \approx 58$-$73$ K of the intra-gap manybody resonance on $E_F$ (see inset in fig.10b), and, finally, to estimate the effective mass $m^*(T \approx 100K) \approx 24\, m_0$ and localization radii of these many-body states $a^*_{p1} \approx 5$ Å (range I, 120-300 K) and $a^*_{p2} \approx 9$ Å (range II, 50-120 K). Because of the spatial extension of the complexes $a^*_{p1} \approx 5$ Å is about equal to the shortest distance (~5.3 Å) between the $R^{3+}$ ions in the $RB_{12}$ lattice, it provides a strong arguments in favor of the conclusion of Ref. 58 concerning formation of $R^{3+}$-$R^{3+}$ dimmers and an appearance of a short range RE-based vibration clusters in the matrix of the dodecaboride compounds. A similar conclusion was made in Ref.67 where the Yb-Yb pairs' existence was deduced from the results of electron spin resonance experiments in the narrow-gap semiconductor $YbB_{12}$. Moreover, the dipole-dipole splitting of the resonance lines permitted estimation[66] of the distance $r_{p2}=9.1$ Å between the interacting ytterbium pairs in $YbB_{12}$ which is about equal to the $a_{p2}$ value in the $Tm_{1-x}Yb_xB_{12}$ compounds.

Taking into account that the localization radius of the intra-gap states $a^*_{p2} \approx 8$-$9$ Å in the range II exceeds the lattice constant ($a \sim 7.5$ Å) and the size increases with Yb concentration $x$ in the $Tm_{1-x}Yb_xB_{12}$ series, the coherent regime of the charge transport may be attributed to a percolation in the charge carriers' motion through the manybody states' network in the $RB_{12}$ matrix. From this point of view a quite natural interpretation may be suggested to explain the origin of the unusual dependence Eq.(7) for the Hall coefficient behavior in the coherent state. Indeed, on the metallic side of percolation limit the magnetic system of the intra-gap manybody states which is arranged on the small size magnetic clusters of RE-ions should be considered as a system with topologically nontrivial configuration of the core spins. As a result, the appearance of Berry phase effect produces the internal magnetic field $H_{int}= <h_z> \sim (1/k_BT)\, exp(-T_0/T)$ and Hall coefficient is modified in according with the prediction [35-36]. It is worth noting also that simultaneously with a drastic decrease of the Hall coefficient at low temperatures and the $R_H(T)$ sign inversion observed in the coherent regime (interval III), a nearly linear temperature dependence of Seebeck coefficient was established for $Tm_{1-x}Yb_xB_{12}$ solid solutions at temperatures $T<30$ K [10]. A variation of the slope of the Mott-type diffusive thermopower $S(T)$ in the metallic coherent state allowed the authors [10] to estimate approximately as an order of magnitude the renormalization effect in DOS near the Fermi energy $E_F$ when the manybody resonance appears with $x$ increase in the series of the $Tm_{1-x}Yb_xB_{12}$ solid solutions (see also inset in fig.10b).

It should be stressed that the local on-site *4f-5d* spin fluctuations play the dominant role in the charge transport of $Tm_{1-x}Yb_xB_{12}$ dodecaborides. Indeed, when $n_{4f}$ increases from $HoB_{12}$ to $LuB_{12}$ throughout the range $10 \leq n_{4f} \leq 14$, one should expect a monotonous *increase* in $\mu_H(n_{4f})$, which follows from a decrease of the de Gennes factor $(g_J-1)^2 J(J+1)$ characterizing the magnetic scattering intensity in the $RB_{12}$ series. Instead of that, below the QCP in the range $10 \leq n_{4f} \leq 12.3$ a *decrease* in $\mu_H(n_{4f})$ is observed (fig.10a), and the mobility minimum near $x = 0.3–0.5$ in the $Tm_{1-x}Yb_xB_{12}$ corresponds to the AF–P transition with the quantum critical point near $x_C \approx 0.3$. Further on, substitution of Yb for Tm in the range $x \geq 0.5$ ($n_{4f} \geq 12.5$ in fig.10a) leads to an increase of $\mu_H(x)$, while a monotonous and sharp growth of the resistivity $\rho(x)$ (fig.1) and Hall coefficient $R_H(x)$ (fig.10b) at low temperatures is observed in the range $0 < x < 1$ (the values of $\rho(T)$ and $R_H(T)$ in $YbB_{12}$ reach 10 $\Omega\cdot$cm and 88 cm$^3$/C respectively [11,69-70]). The metal–insulator transition is accompanied by a rather low mobility values $\mu_H(T) \approx 27$ cm$^2$/(V s) in $YbB_{12}$ that are obviously caused by carriers scattering on the strong spin and charge *4f–5d* fluctuations. Then, as the narrow-gap insulator $YbB_{12}$ ($n_{4f} \approx 13$) changes to the nonmagnetic superconducting metal $LuB_{12}$ ($n_{4f} = 14$), a drastic increase in the Hall mobility is observed up to $\mu_H(T) \approx 2600$ cm$^2$/(V s) (see fig. 10a).

When comparing of the two types' systems – the metallic heavy fermion compounds near the AF QCP, from one hand, and $Tm_{1-x}Yb_xB_{12}$ solid solutions with QCP and MIT, from another it is worth noting the results of inelastic neutron scattering studies obtained by Stockert et al. in [50,68] for $CeCu_{5.9}Au_{0.1}$ and K. S. Nemkovski et al. for $YbB_{12}$ [27]. The authors [50,68] have concluded that the dimension of the magnetic excitation spectrum decreases near the QCP in CeCu$_{5.9}$Au$_{0.1}$ and its features coincide with the magnetic reflections registered in the AF state of antiferromagnets CeCu$_{5.8}$Au$_{0.2}$ and CeCu$_{5.7}$Au$_{0.3}$. A similar quasi-two-dimensional (2D) character of the spin fluctuation spectrum was revealed in [27] for YbB$_{12}$. Indeed, the M1 peak in the magnetic excitation spectrum of YbB$_{12}$ was ascribed to AF correlations at the wave vector (½ ½ ½). Assuming the intensity distribution to reflect the range of AF interactions between the Yb moments the authors[27] have estimated correlation lengths perpendicular and parallel to the above direction, corresponding, respectively, to couplings within ($\xi_\parallel$) or between ($\xi_\perp$) the (001) planes. The values $\xi_\parallel$=5.4±1.4 Å and $\xi_\perp$=3.4±1.1 Å obtained in Ref. 27 confirm the 2D-character anisotropy. Moreover, from our view, it allows one to suppose of some manybody states' network formation with the most effective Yb-Yb pairs' interaction in YbB$_{12}$.

An interpretation of these 2D effects in CeCu$_{5.9}$Au$_{0.1}$ was done in [6] in terms of a two types fluctuations- local on-site *4f-5d* fluctuations arising from instability of the RE-ion configuration and the long-wavelength fluctuations due to the development of antiferromagnetic instability in vicinity of quantum phase transition. According to the approach [6] these first type fast local *4f-5d* spin fluctuations lead to a renormalization of the effective mass of charge carriers through the formation of spin polarons (resonance at $E_F$) in the heavy fermion paramagnets. The appearance of the long-wavelength fluctuations near the QCP at $T_N$=0 in addition to the on-site ones is equivalent to "switching on" an interaction between the spin polaron complexes, leading to the formation of a network of interconnected heavy fermion states. As a result, the character of the quasiparticle interactions changes abruptly at the QCP (non-Fermi-liquid behavior)[53-56], and one observes the lowering of the dimensionality of the spin fluctuations [49,50], unusual scaling of the dynamic magnetic susceptibility[7-8] etc. In our opinion, the same mechanism is responsible for the manybody states' network formation in the matrix of Tm$_{1-x}$Yb$_x$B$_{12}$ solid solutions. However, in this case on the way from the metal with strong *4f-5d* spin fluctuations (TmB$_{12}$) to the intermediate valence narrow-gap semiconductor (YbB$_{12}$) the local on-site fluctuations increase dramatically, and in addition to the AF QCP at $x_C$~0.3 (AF instability) both the metal-insulator transition (fig.10b) and the vibration instability are developed with a cage-glass state formation at temperatures below $T^*\approx 70K$[58]. In our view, the even component in the Hall effect observed in this study, which appears and increases drastically above the QCP (fig. 10b) is also due to the formation of a structure of interconnected spin polaron states in the conduction band of the $Tm_{1-x}Yb_xB_{12}$ compounds.

## 4. CONCLUSION.

The present study provides the results of the detailed measurements of Hall effect in the substitutional solid solutions $Tm_{1-x}Yb_xB_{12}$ with $x<0.75$ and it has enabled to separate and classify the contributions in the charge transport in these strongly correlated electron systems. The comprehensive analysis of the angular, temperature and field dependences of the Hall resistivity was carried out for compounds close to antiferromagnet-paramagnet and metal-insulator transitions in a wide vicinity of a quantum critical point at $x_C\sim 0.3$ at temperatures 1.9-300K in magnetic field up to 8 T. It was found for the first time the appearance of the second harmonic contribution on the angular dependences of the Hall resistivity of the $Tm_{1-x}Yb_xB_{12}$ dodecaborides nearby QCP and its' enhancement with the concentration growth in the interval $x \geq x_C$. A pronounced negative minimum on the Hall coefficient temperature dependence for $x \geq x_C$ and sign inversion of $R_H(T)$ at helium temperatures for $Tm_{1-x}Yb_xB_{12}$ compounds with $x \geq 0.5$ was revealed and the $R_H(T)$ decrease at $T< 30K$ was identified as a feature of the coherent regime of the charge transport. It was shown that the external magnetic field leads to an enhancement of the second harmonic contribution in the Hall resistivity and it suppresses the coherent regime of charge carriers' spin-flip scattering on the magnetic moments of the rare-earth ions. For the $Tm_{1-x}Yb_xB_{12}$ compounds with $x \geq 0.5$ in the intervals 120-300 K and 50-120 K we have found an activation type dependence of the Hall coefficient with the activation energies $E_g \approx 17.8$ meV and $E_a/k_B \approx 58$-73 K and the microscopic parameters: effective masses $m^* \sim 20\ m_0$ and localization radii of the heavy fermion many-body states $a^*_{p1} \approx 5$ Å (120-300 K) and $a^*_{p2} \approx 9$ Å (50-120 K). In the framework of the spin-polaron approach the appearance of these odd and even anomalous components $R_H(T,H)$ and $\rho_{H2}(T,H)$ in the Hall effect of the $Tm_{1-x}Yb_xB_{12}$ series is discussed in terms of the interference effects between local *4f-5d* and long-range spin fluctuations, leading to the formation of a magnetic structure (network) of the interconnected manybody complexes in the dielectric matrix of these compounds.

We are grateful to G.E. Grechnev, A.V. Kuznetsov and D.A. Parshin for stimulating discussions. This work was supported by the Branch of Physical Sciences, Russian Academy of Sciences (program "Strongly Correlated Electrons in Metals, Semiconductors, and Magnetic Materials"), the Russian Foundation for Basic Research (project no. 10-02-00998-a) and by Russian State Programme "Human resources of Innovative Russia". N.E.S., A.V.B. and V.V.G. are grateful to Institute for Nanoscale Physics and Chemistry of KULeuven for hospitality and support during their short research visits.

**Figure captions.**

**Fig. 1.** The temperature dependences of resistivity for $Tm_{1-x}Yb_xB_{12}$ compounds. The inset shows the schema of the Hall effect measurements (see text).

**Fig. 2.** The angular dependences of the Hall resistivity $\rho_H(\varphi)$ for $Tm_{0.46}Yb_{0.54}B_{12}$, recorded in the magnetic field $\mu_0H=1.5$ T at various temperatures in the range 2÷35 K. For all curves $\rho_H(\varphi)$ the decomposition on the odd and even harmonics is presented (see Eq.(1) and text).

**Fig. 3.** The angular dependences of the Hall resistivity $\rho_H(\varphi)$ for $Tm_{0.28}Yb_{0.72}B_{12}$, recorded at helium temperature $T=2.1$ K in magnetic fields $\mu_0H\leq 8$ T. For all curves $\rho_H(\varphi)$ the decomposition on the odd and even harmonics is presented (see Eq.(1) and text).

**Fig. 4.** The temperature dependences of Hall coefficient $R_H(T,H_0)=\rho_{H1}(H,T)/H$ for solid solutions $Tm_{1-x}Yb_xB_{12}$ with $x=0.31$, 0.54 and 0.72 in magnetic field $\mu_0H=1.5$ T, 7 T and 8 T. The $R_H(T,H_0)$ dependence for $TmB_{12}$ ($x=0$) compound in magnetic field $\mu_0H=0.37$ T is shown for comparison.

**Fig.5.** The temperature dependences of the second harmonic component $\rho_{H2}(H,T)$ in Hall resistivity (a) for $\mu_0H=1.5$ T in the hyperbolic plot (see Eq.(8)) and (b) for $\mu_0H=7$-8 T in the double logarithmic coordinates.

**Fig. 6.** Magnetic field dependences of the Hall coefficient $R_H(H)$ at helium temperatures $T_0=4.2$ K (closed symbols) and $T_0=2.1$ K (open symbols) for $Tm_{1-x}Yb_xB_{12}$ compounds with $x=0.31$, 0.54 and 0.72.

**Fig. 7.** Magnetic field dependences of the second harmonic component $\rho_{H2}(H,T)$ in Hall resistivity at helium temperatures $T_0=4.2$ K (closed symbols) and $T_0=2.1$ K (open symbols) for $Tm_{1-x}Yb_xB_{12}$ compounds with $x=0.31$, 0.54 and 0.72.

**Fig.8.** An activation type dependences of the reduced charge carriers' concentration $\nu=(R_HeN_{4f})^{-1}$ for $Tm_{1-x}Yb_xB_{12}$ compounds with $x=0.31$, 0.54 and 0.72 in the different temperature ranges I-III (see text).

**Fig.9.** Hall mobility $\mu_H(T)=R_H(T)/\rho(T)$ temperature dependences in the series of $Tm_{1-x}Yb_xB_{12}$ compounds.

**Fig.10.** (a) Hall mobility $\mu_H(n_{4f})$ at helium temperature (for $HoB_{12}$ ($n_{4f}$=10), $ErB_{12}$ ($n_{4f}$=11) and $LuB_{12}$ ($n_{4f}$=14) the data taken from Refs.43,57 and for $YbB_{12}$ ($n_{4f}$~13.05)-from Ref. 70) and the Neel temperature $T_N(n_{4f})$ vs the filling of the 4*f* shell of the rare earth ions in the $RB_{12}$ compounds; AF and P denote the antiferromagnetic and paramagnetic phases, and M, I and SC – metal, insulator and superconductor respectively, QCP- the quantum critical point, (b) Hall coefficient $R_H(n_{4f})$ and even harmonic contribution $\rho_{H2}(n_{4f})$ for $\mu_0H$ =1.5 T and for $\mu_0H$ =8 T, respectively. The inset shows the DOS renormalization effect in the series of $Tm_{1-x}Yb_xB_{12}$ compounds.

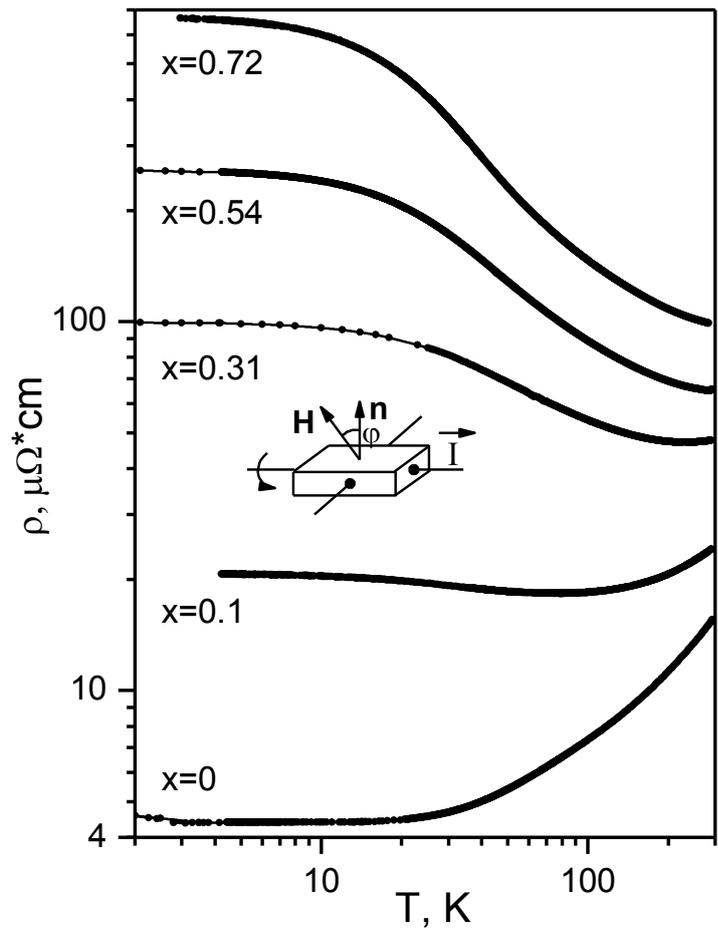

Fig. 1

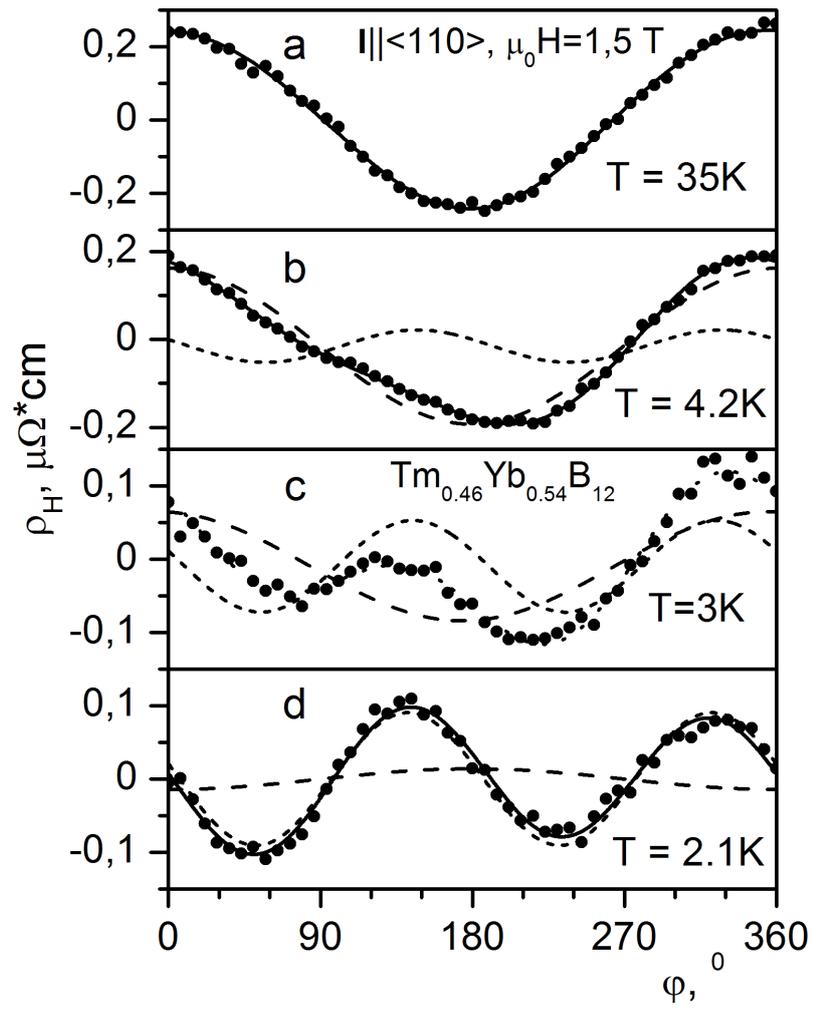

Fig. 2

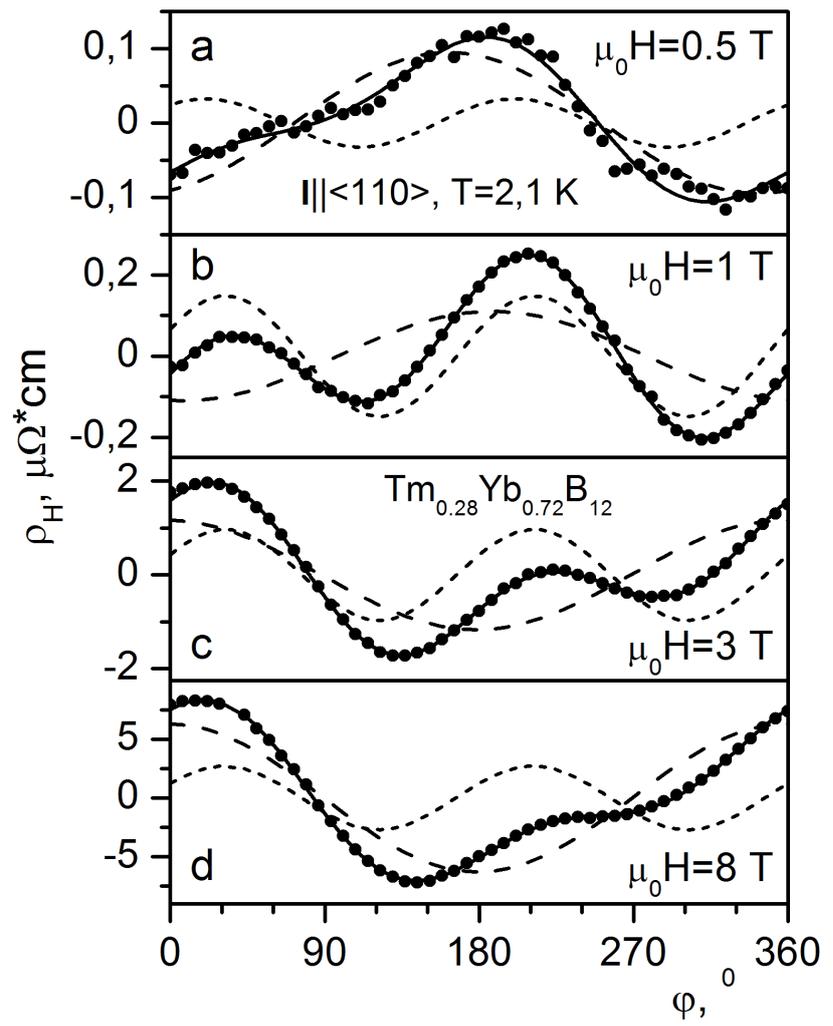

Fig. 3

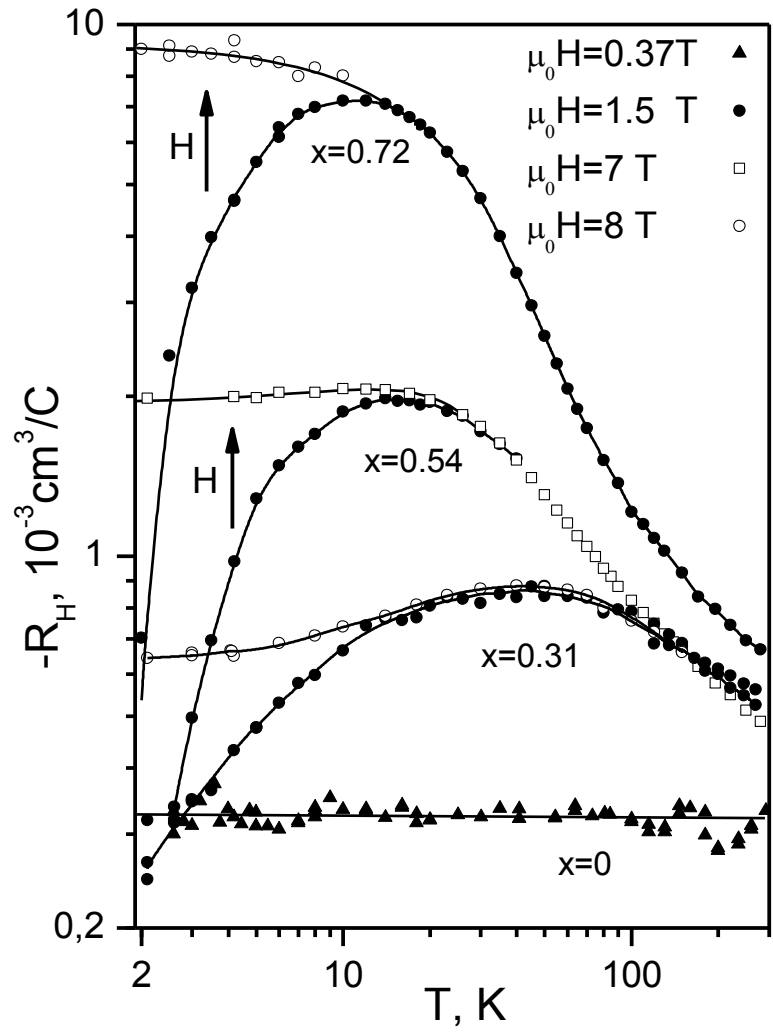

Fig.4

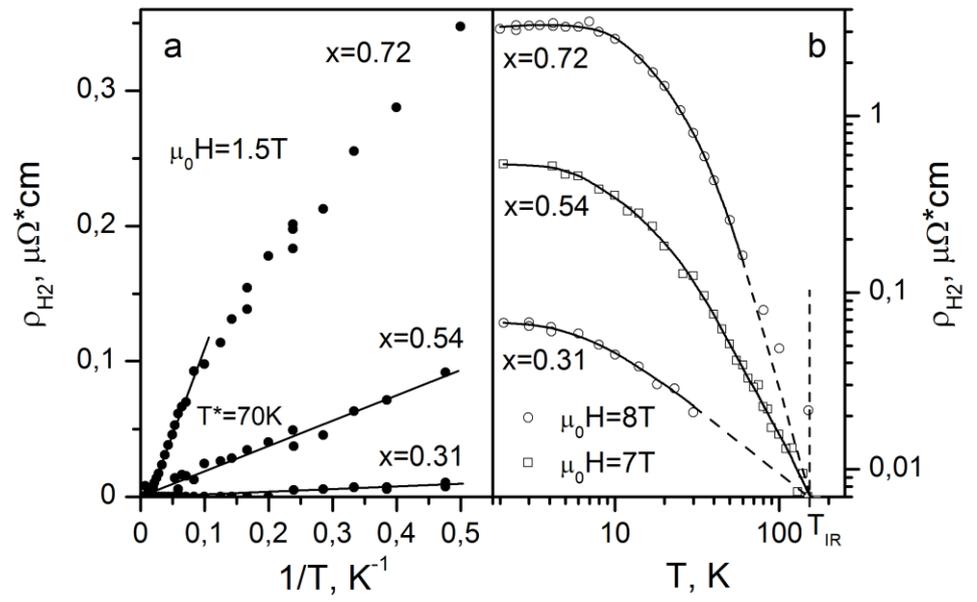

Fig.5

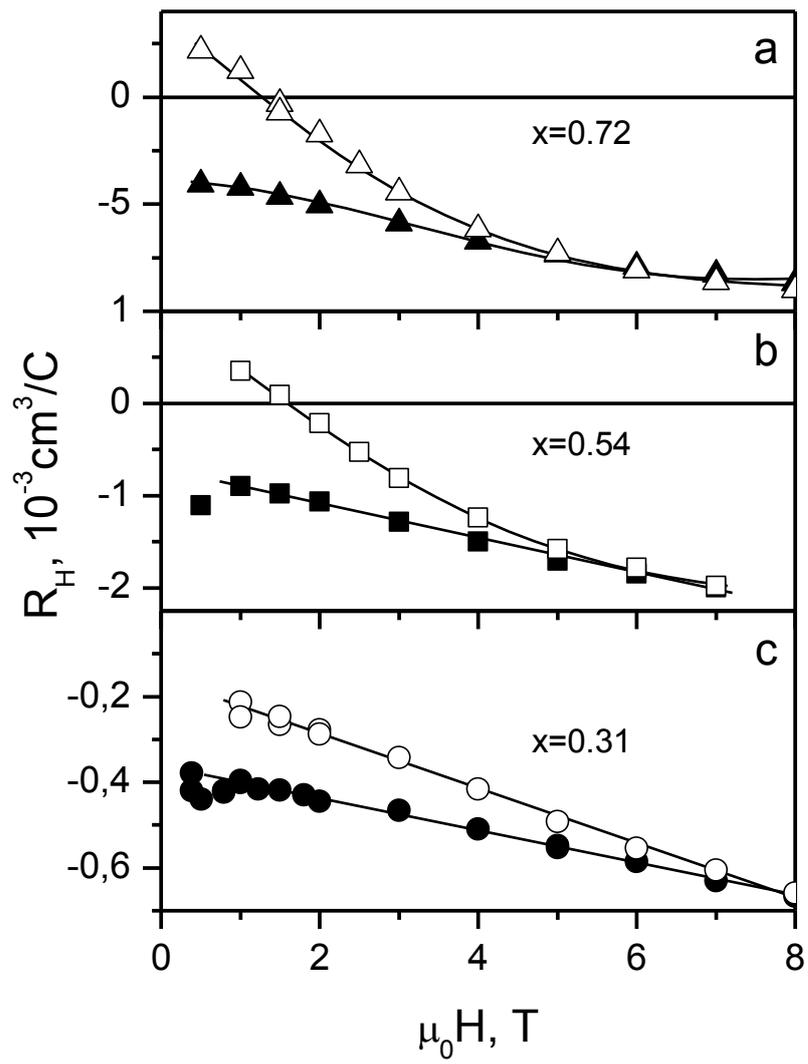

Fig.6

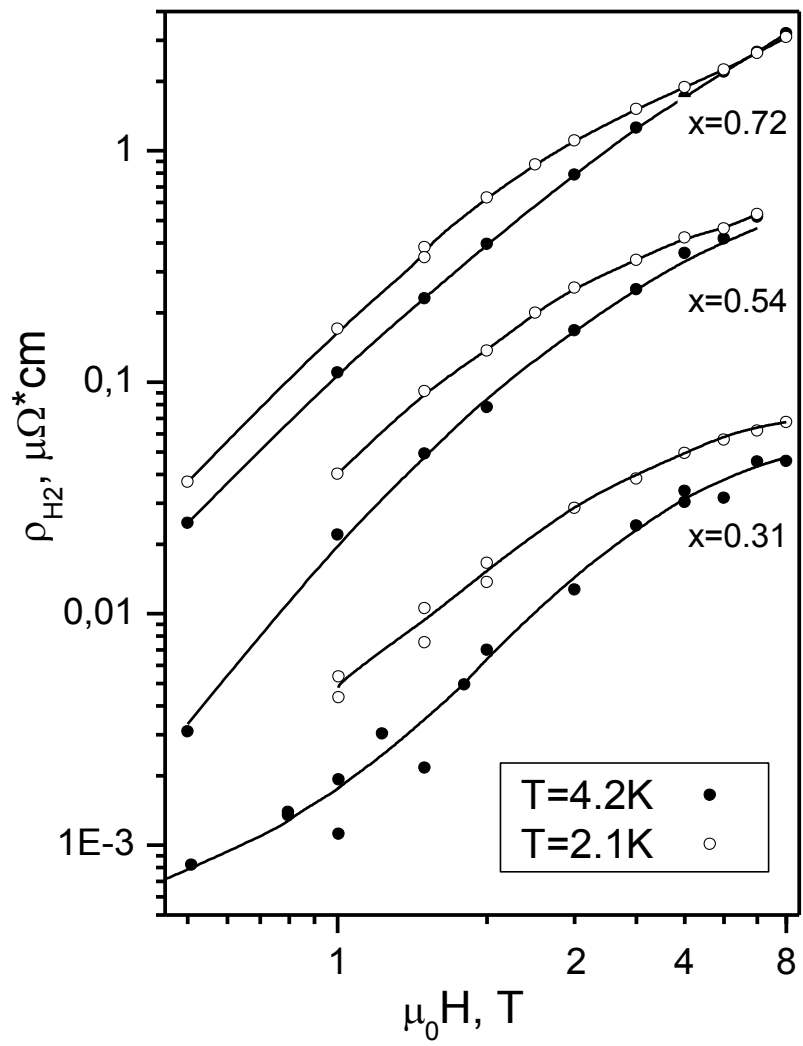

Fig.7

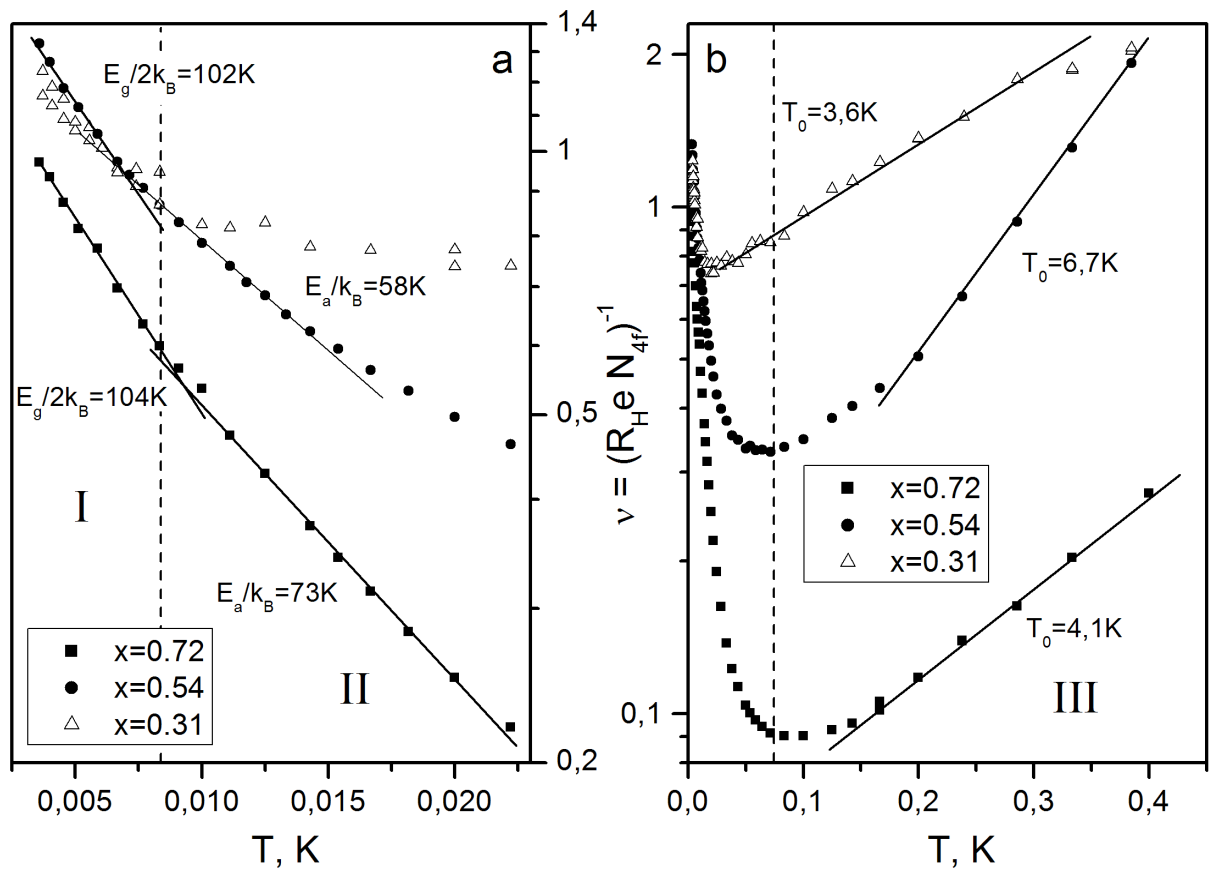

Fig.8.

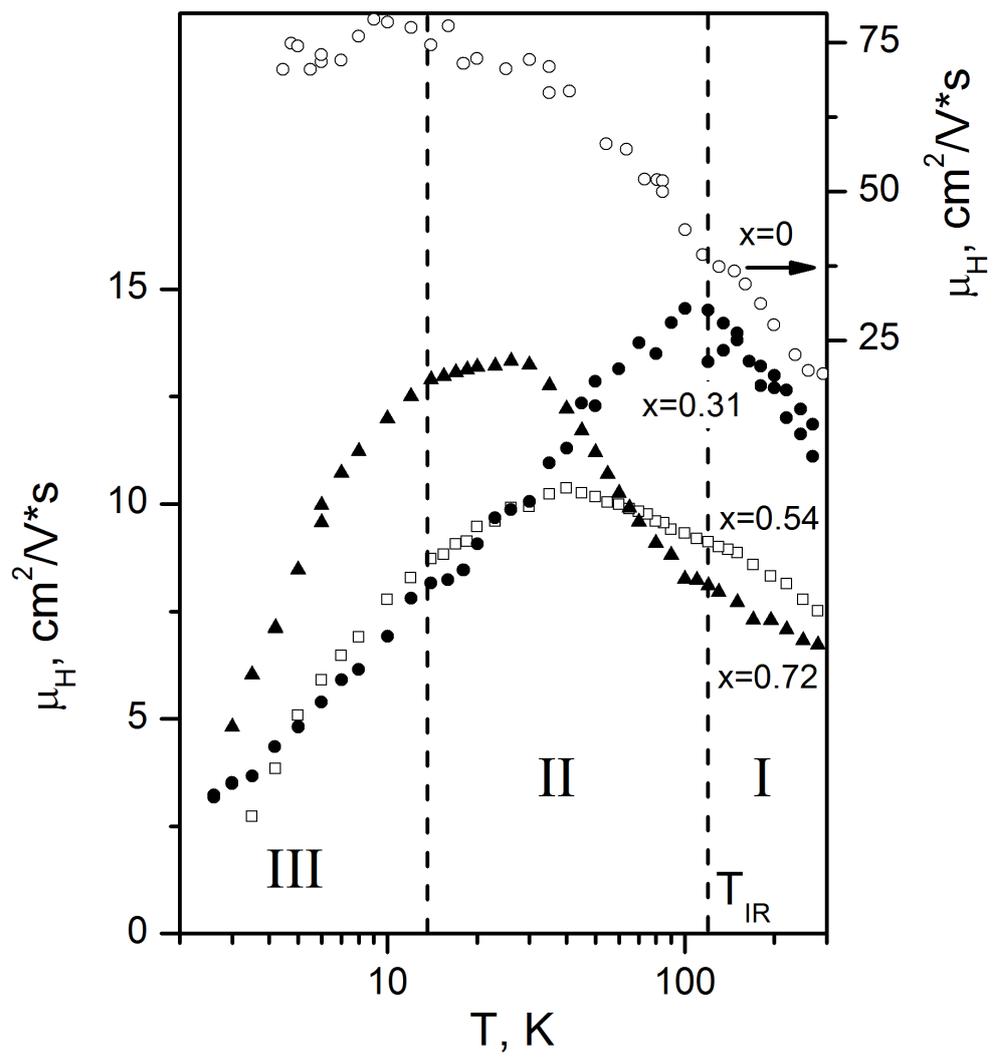

Fig.9

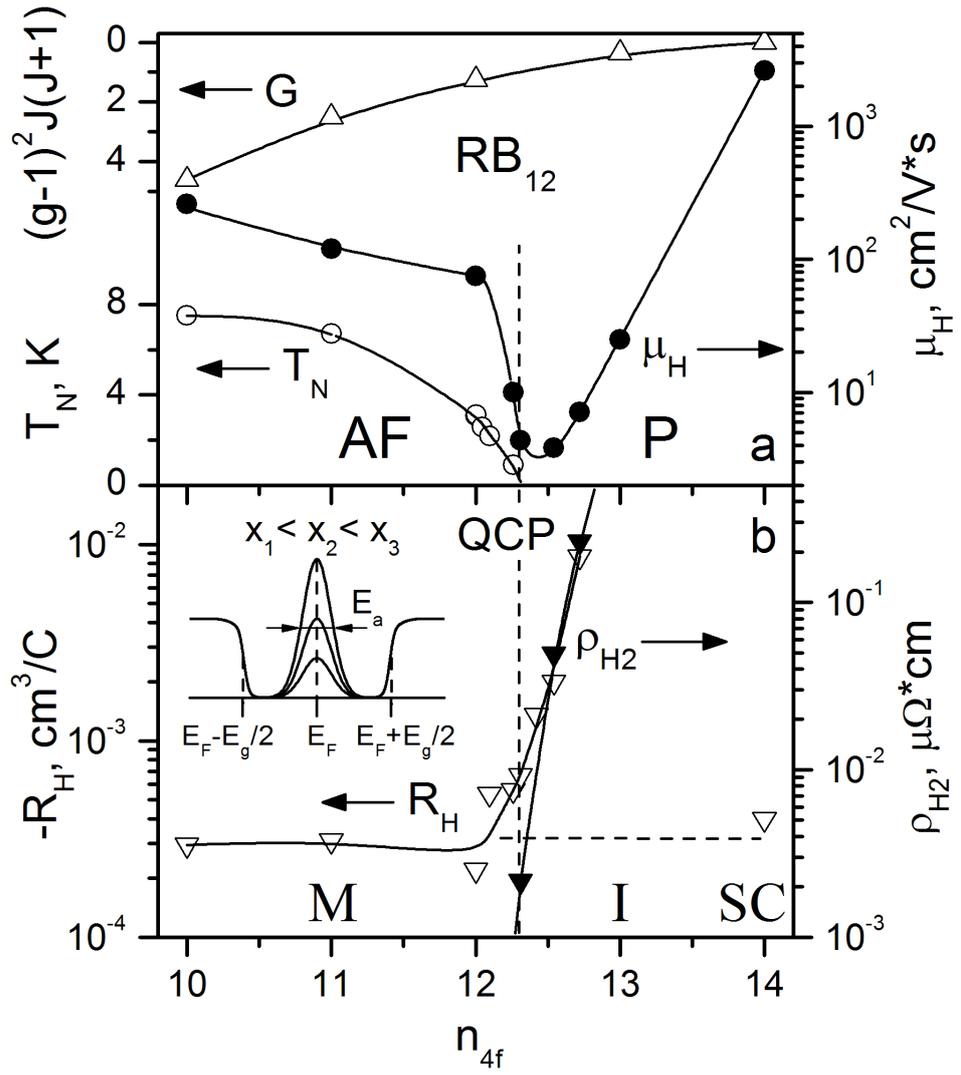

Fig.10